\documentclass{article}
\usepackage{graphicx}
\usepackage{hyperref} 
\pdfoutput=1 
\begin{document}
\begin{center}
{\Large{\sc The short life of a drop}}\\
\bigskip
Guillermo Hern\'andez-Cruz$^1$, Minerva Vargas$^2$, Heberto P\'erez$^1$\\
J. Arturo Pimentel$^3$, Gabriel Corkidi$^3$  and Eduardo Ramos$^1$\\
\bigskip
$^1$Center for Energy Research, Universidad Nacional Aut\'onoma de M\'exico\\
$^2$ Instituto Tecnol\'ogico de Zacatepec\\
$^3$ Biotechnology Institute, Universidad Nacional Aut\'onoma de M\'exico
\end{center}
\bigskip
\begin{abstract}
This is the companion paper of the Fluid Dynamics Video "The short life of a drop" where it is argued that the geometry of the sediment of a drop of water with particles in suspension can be correlated with the dynamics of the fluid inside the drop during the drying process.
\end{abstract}
\bigskip
We study the dynamics of the flow inside a sessile drop of  water with a suspension of polystyrene spheres of 1$\mu$m in diameter. According to our observations it is possible to correlate the motion of the fluid inside the drop with the pattern of sediments. The initial volume of the drops is 1$\mu$l  and the motion is recorded with a micro PIV equipment that comprises an Inverted Optical Microscope (Olympus IX71) and a  video camera (Optronis CR5000x2) that captures 512 $\times$ 512 images at a rate of 60 fps. The microscope has an amplification of 4X and the objective has a nominal depth of field of 40 $\mu$m. The drop sits on a soda lime glass substrate that is kept at 20 $^\circ$C. The initial footprint of the drop is 2 mm in diameter and the suspension has 5.67 $\times 10^6$ spheres per $\mu$l of suspension. The evaporation process lasts typically several minutes with the exact duration depending on the relative humidity in the room. The evaporation process can be divided in four parts. In the first, the drop takes the shape of a spherical cap and  the velocity of the fluid inside the drop is very slow and difficult to resolve but particles emigrate steadily towards the contact line and build up to start forming the coffee ring. In the second part, the drop is almost flat and the velocity field inside the drop is mostly radial to compensate for the preferential evaporation near the contact line. In this part, the radial velocity near the edge diverges as the thickness of the liquid layer is reduced and the rapid accumulation of the particles near the contact line result in a quick thickening of the coffee ring. In these two first stages, the edge of the drop remains pinned and can be well approximated by a circle. The third part of the process starts when the liquid film is sufficiently thin and rips off from the coffee ring and a complicated interaction, of surface tension (which pulls the liquid to reduce the area) and evaporation, determine the internal flow.  In the initial stages of this part  a transient formation of a secondary coffee ring slows down the motion of the retracting contact line forming relatively thick sediment segments with the shape of short circular arcs. Subsequently, the geometry of the outer boundary of the drop  becomes unstable and small perturbations modify its local curvature to generate a time dependent non-circular edge.  During the fourth part of the process, local irregular features of the drop edge are smoothed out by surface tension. This effect also generates fast inward radial flows which sweep particles forming thin radial segments of sediments. This situation prevails until the liquid totally evaporates.
\end{document}